\documentclass[12pt]{article}
\usepackage{latexsym,amsfonts}

\usepackage{color}
\usepackage{amsmath}
\usepackage{amssymb}
\usepackage{epsfig}
\usepackage{bm}

\begin{document}
\title{\bf \LARGE $\phi^6$ kink scattering}
\author{{Stephen W. Goatham{\thanks{E-mail: {\tt
	swgoatham@hotmail.com}}}}\hspace{1.5mm}\\ \\[5pt]
{\normalsize {\sl School of Mathematics, Statistics and Actuarial
Science}}\\
{\normalsize {\sl University of Kent,
Canterbury CT2 7NF, United Kingdom}}
}

\date{ \today}
\maketitle

\begin{abstract}
In this paper, we investigate the  scattering of two $\phi^6$ kinks and derive the real dynamics by solving the appropriate field equation numerically employing a Runga-Kutta method. We also use a collective coordinate approximation to find approximate dynamics, with the objective being to compare the approximate motion to the real dynamics in order to test the validity of this kind of approximation, which is used extensively in the study of solitons.
\end{abstract}

\section{Introduction}
Topological solitons are smooth, stable, particle-like solutions to non-linear wave equations that occur in field theories \cite{Manton:2004tk}.  These include solitons such as monopoles in (3+1) dimensions and vortices and lumps in (2+1) dimensions.  However, although some of the physically interesting theories with solitons may be exactly solvable in the static case (through the reduction of second order to first order differential equations via a Bogomolny argument \cite{Bogomolny}), they are usually not integrable upon the introduction of time dependence.  This makes it difficult to study soliton dynamics and usually one must make use of approximations or numerical simulations.  Here, we present both of these approaches, this time in the context of $\phi^6$ kinks. Our aim is to use one method to test the other. \footnote{Note that kink-antikink scattering in the $\phi^6$ model has also been investigated recently, see \cite{Dorey:2011yw}} Section 2 presents a discussion of $\phi^6$ kinks.  Then section 3 introduces our collective coordinate approximation, with section 3.1 discussing one $\phi^6$ kink and section 3.2 focusing on two kinks.  Section 4 presents dynamics for two kinks.  The paper ends with a conclusion.

\section{$\phi^6$ kinks}
The $\phi^6$ model is a field theory in (1+1) dimensions involving a real field $\phi(t,x).$  The Lagrangian density is given by

\begin{equation}
\label{phi6}
{\cal L}=\frac{1}{2}{\left(\frac{\partial \phi}{\partial t}\right)}^2-\frac{1}{2}{\left(\frac{\partial \phi}{\partial x}\right)}^2-U(\phi),
\end{equation}
\noindent
where

\begin{equation}
U(\phi)=\frac{1}{2}\phi^2{(1-\phi^2)}^2.
\end{equation}
\noindent
Varying (\ref{phi6}) leads to the equation of motion

\begin{equation}
\frac{\partial^2 \phi}{{\partial t}^2}-\frac{\partial^2 \phi}{{\partial x}^2}+\frac{dU}{d\phi}=0.
\end{equation}
\noindent
The Bogomolny equations are
\label{boggeneral}
\begin{equation}
\phi'=\pm \sqrt{2U}.
\end{equation}
\noindent
For the $\phi^6$ model they become

\begin{equation}
\label{phi6bog}
\phi'=\pm \phi(1-\phi^2).
\end{equation}
This ODE can be solved exactly for both signs.  Taking the negative sign, we find a kink (with $\phi(-\infty)=-1$ and  $\phi(\infty)=0$) and an antikink (with $\phi(-\infty)=0$ and  $\phi(\infty)=-1$).  The kink solution is

\begin{equation}
\label{phi6kink1}
\phi_{1}(x)=\frac{-1}{\sqrt{1+\exp{(2(x+a)+\log{3})}}},
\end{equation}
where $a$ has been introduced as a constant of integration.  Note that the factor $\log{3}$ has been included to position the kink at the origin when $a=0.$

Now taking the positive sign in (\ref{phi6bog}), we again find a kink (this time with $\phi(-\infty)=0$ and  $\phi(\infty)=1$) and an antikink (with $\phi(-\infty)=1$ and  $\phi(\infty)=0$).  The kink solution is

\begin{equation}
\label{phi6kink2}
\phi_{2}(x)=\frac{1}{\sqrt{1+\exp{(-2(x-a)+\log{3})}}},
\end{equation}
\noindent
where $a$ is again an arbitrary constant and $\log{3}$ is included.

In both (\ref{phi6kink1}) and (\ref{phi6kink2}), the constant $a$ represents the position or modulus of the soliton.  $\phi_{1}$ and $\phi_{2}$ represent kinks positioned at $-a$ and $a$, respectively.  Note we have the symmetry

\begin{equation}
\phi_{1}(-x)=-\phi_{2}(x).
\end{equation}
\noindent
We can demonstrate that $-a$ and $a$ are the moduli of the kinks by considering the energy density

\begin{equation}
\label{energydensity}
{\cal E}=\frac{1}{2}\phi'^2+U=\phi'^2.
\end{equation}
\noindent
Here $\phi$ is a static kink field.  (\ref{energydensity}) implies

\begin{equation}
\frac{d{\cal E}}{dx}=2\phi'\phi''.
\end{equation}
\noindent
Consider (\ref{phi6kink1}).  When $x=-a,$ $\phi_{1}'\ne0.$  Hence, for this point to represent the maximum of the energy density and therefore the kink position we need $\phi_{1}''(-a)=0.$  Differentiating (\ref{phi6kink1}) we see that this is indeed the case.  Similarly we find $\phi_{2}''(a)=0.$

The dynamics of a kink given by (\ref{phi6kink1}) or (\ref{phi6kink2}) can be found by a simple Lorentz boost, e.g.

\begin{equation}
\phi_{1}(t,x)=\frac{-1}{\sqrt{1+\exp{(2(\gamma(x-vt+a+\frac{\log{3}}{2})))}}},
\end{equation}
\noindent
where $v$ is, as usual, the kink speed (with $-1<v<1$) and $\gamma=\frac{1}{\sqrt{1-v^2}}$ is the Lorentz factor.  Note that while the dynamics of one kink, in this model, can be found by direct integration, those of multi-kinks cannot.  This is a result of the fact that the $\phi^6$ model is not an example of an integrable system and therefore, unlike for the well-studied sine-Gordon model, solution generating techniques such as the B\"acklund transformation are not applicable.

\section{The collective coordinate approximation}
\subsection{One kink}
In \cite{Manton:1981mp}, a moduli space approximation was introduced.  This kind of approximation was originally implemented for a system of solitons for which there are no static forces.  In the case of monopoles it is possible to have multi-soliton configurations that are static with no forces between solitons (critical coupling) because of a precise balancing between a magnetic force and a scalar interaction (due to a Higgs field).  This leads to geodesic trajectories of multi-solitons on the space of static configurations (the moduli space).  For kinks, of course, only a single kink can be static and the calculation is simple.  We demonstrate this for a $\phi^6$ kink.

A static kink is given by (\ref{phi6kink2}).  Here the parameter or modulus is a real number $a$ and hence the moduli space is simply $\mathbb R.$  The moduli space approximation is to consider a flow on this manifold, i.e. to make $a$ time-dependent, and to then substitute this dynamical field into (\ref{phi6}).  Integrating over all space gives the dynamics

\begin{equation}
L=\frac{1}{8}{\dot a}^2-\frac{1}{4},
\end{equation}
\noindent
hence motion at a constant speed.  Generally, when there are no static forces, there is geodesic motion on the moduli space.

\subsection{Two kinks}
\begin{figure}[!htb]
\begin{center}
\begin{includegraphics}[width=12cm]{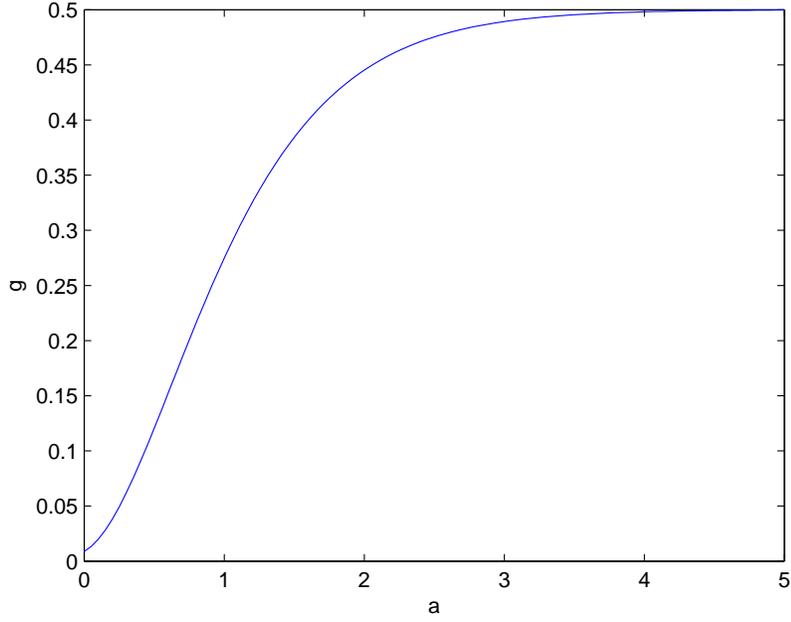}\end{includegraphics}
\caption{The metric $g$ as a function of kink position $a$.\label{metric}}
\end{center}
\end{figure}

\begin{figure}[!htb]
\begin{center}
\begin{includegraphics}[width=12cm]{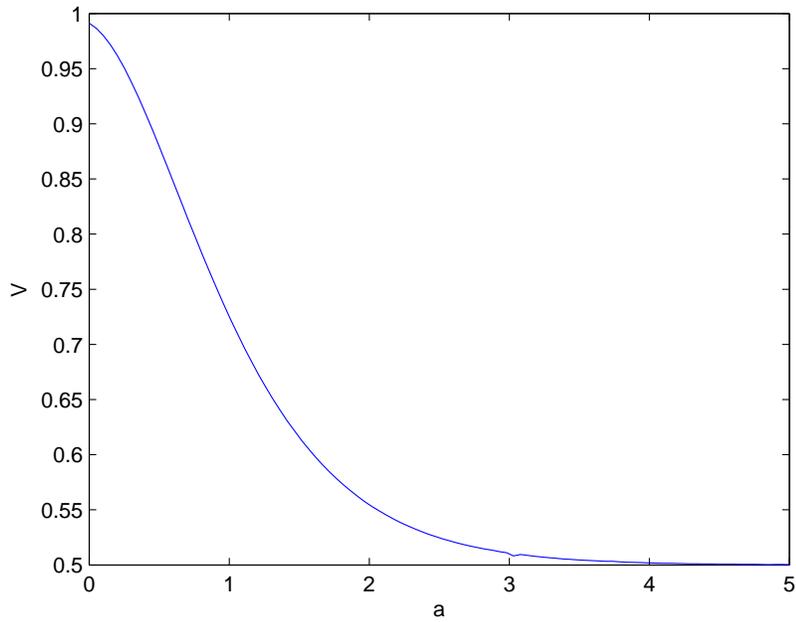}\end{includegraphics}
\caption{The potential $V$ as a function of kink position $a$.\label{Potential}}
\end{center}
\end{figure}

When there are static forces, the moduli space approximation becomes harder to implement.  We consider multi-solitons that experience static forces.  Here, it is possible to make progress by taking 1-soliton configurations that are static and patching these into one field then performing a flow.  When the solitons are far apart the field will be accurate.  It is then hoped that when the solitons approach and interact, the approximation will still be good.  This approach has been applied for solitonic structures such as lumps, vortices and monopoles, as well as kinks.  See, for example, \cite{Atiyah,Gibbons:1986df} for a study of classical and quantum monopole scattering, \cite{Ruback:1988ba,Samols:1991ne} for vortex scattering, and \cite{Ward:1985ij} for classical lump scattering.  There is confidence in the validity of such a method. To provide further confidence we can perform ``tests'' on one-dimensional models.  Such a test was performed on the scattering of two sine-Gordon kinks by Sutcliffe \cite{Sutcliffe:1993wc}.  Here we show how to perform such a test on the $\phi^6$ model. We choose to obtain a 2-kink field by patching together two 1-kink solutions in the following way

\begin{equation}
\label{kinkansatz}
\phi=\frac{-1}{\sqrt{1+\exp{(2(x+a)+\log{3})}}}+\frac{1}{\sqrt{1+\exp{(-2(x-a)+\log{3})}}},
\end{equation}
\noindent
where $a\in \mathbb R^{+}$ is the collective coordinate, so that ${\cal M}$ is one dimensional.  To perform the truncation to a one dimensional system we substitute the ansatz (\ref{kinkansatz}) into the Lagrangian density (\ref{phi6}), and integrate over all $x.$  This leads to the Lagrangian

\begin{equation}
\label{langragianphi6}
L=\frac{1}{2}g(a){\dot a}^2-V(a).
\end{equation}
\noindent
Here $g(a)$ is the metric on ${\cal M}$

\begin{equation}
g(a)=\frac{1}{2}-18\int_{-\infty}^{\infty}\frac{e^{4a}dx}{(1+3e^{-2(x-a)})^{\frac{3}{2}}(1+3e^{2(x+a)})^{\frac{3}{2}}}
\end{equation}
\noindent
and $V(a)$ is the potential

\begin{equation}
V(a)=\int_{-\infty}^{\infty}(9V_{1}^2+V_{2}^2(1-V_{2}^2)^2)dx,
\end{equation}
\noindent
where

\begin{equation}
V_{1}=\left(\frac{e^{2(x+a)}}{(1+3e^{2(x+a)})^{\frac{3}{2}}}+\frac{e^{-2(x-a)}}{(1+3e^{-2(x-a)})^{\frac{3}{2}}}\right)
\end{equation}
\noindent
and

\begin{equation}
V_{2}=-\frac{1}{\sqrt{1+3e^{2(x+a)}}}+\frac{1}{\sqrt{1+3e^{-2(x-a)}}}.
\end{equation}
\noindent
In figures \ref{metric} and \ref{Potential},  we plot the metric and potential.  We see that as $a\rightarrow \infty$,  the metric and potential both tend to $0.5.$  We expect this since in this limit there is negligible interaction between the kinks and there will therefore be geodesic motion on ${\cal M}$ with the mass of the system being double the mass of a $\phi^6$ kink.  Indeed the asymptotic interaction energy of two $\phi^6$ kinks is

\begin{equation}
E=2\exp{(-R)},
\end{equation}
\noindent
where $R$ is the kink separation.

In the next section we consider the dynamical system that is given by the Lagrangian (\ref{langragianphi6}) in order to study kink scattering.
\section{The dynamical system}
On varying the Lagrangian (\ref{langragianphi6}) we obtain the field equation

\begin{equation}
\label{eq of motion}
g{\ddot a}+\frac{1}{2}\frac{dg}{da}{\dot a}^2+\frac{dV}{da}=0,
\end{equation}
\noindent
which we can interpret as a particle with position $a(t)$ moving in a potential $V(a)$, with variable  mass $g(a)$. The initial conditions for kink scattering with a velocity  $u$ are $a(t=0)=a_{0}$ and ${\dot a}(t=0)=-u$.  The total energy of the system is

\begin{equation}
E=\frac{1}{2}g(a_{0})u^2+V(a_{0}),
\end{equation}
\noindent
so that the turning point of the motion, $a_{1}$, is given by $V(a_{1})=E$. We can simply solve the equation of motion (\ref{eq of motion}) by quadrature to obtain $a(t)$ implicitly as

\begin{equation}
\label{num in}
t(a)=\int_{a}^{a_{0}}\sqrt{\frac{g(\alpha)}{2(E-V(\alpha))}}d\alpha,
\end{equation}
\noindent
which is valid for $0\le t \le t_{1}$, where $t_{1}$ is the turning time $t_{1}=t(a_{1})$. The position for $t>t_{1}$ is determined by the fact that the motion is symmetric about $t_{1}$, i.e. $a(t-t_{1})=a(t_{1}-t)$.

\begin{figure}[!htb]
\begin{center}
\begin{includegraphics}[width=12cm]{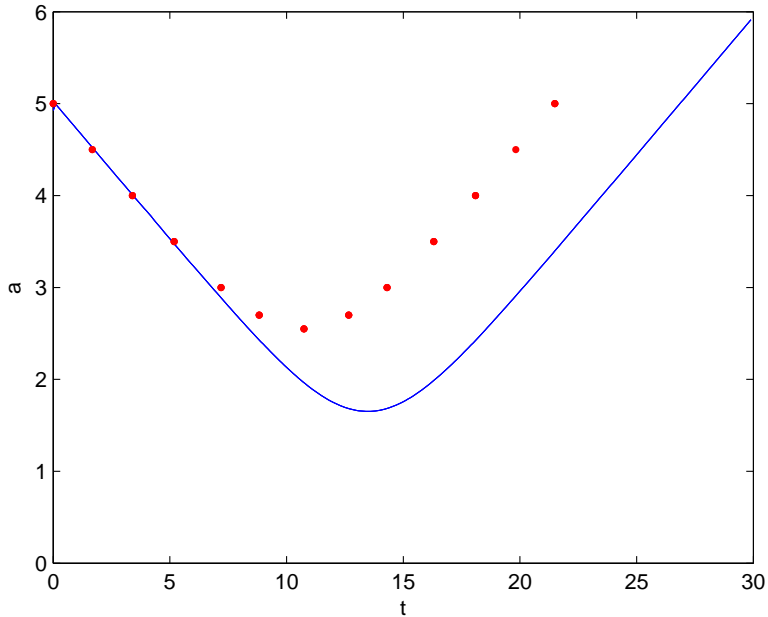}\end{includegraphics}
\caption{The ``real dynamics'' (blue curve) for $a_{0}=5$ and ${\dot a}(t=0)=-0.3$ and the ``approximate dynamics'' (red dots).}
\label{realdyn}
\end{center}
\end{figure}
\begin{figure}[!htb]
\begin{center}
\begin{includegraphics}[width=12cm]{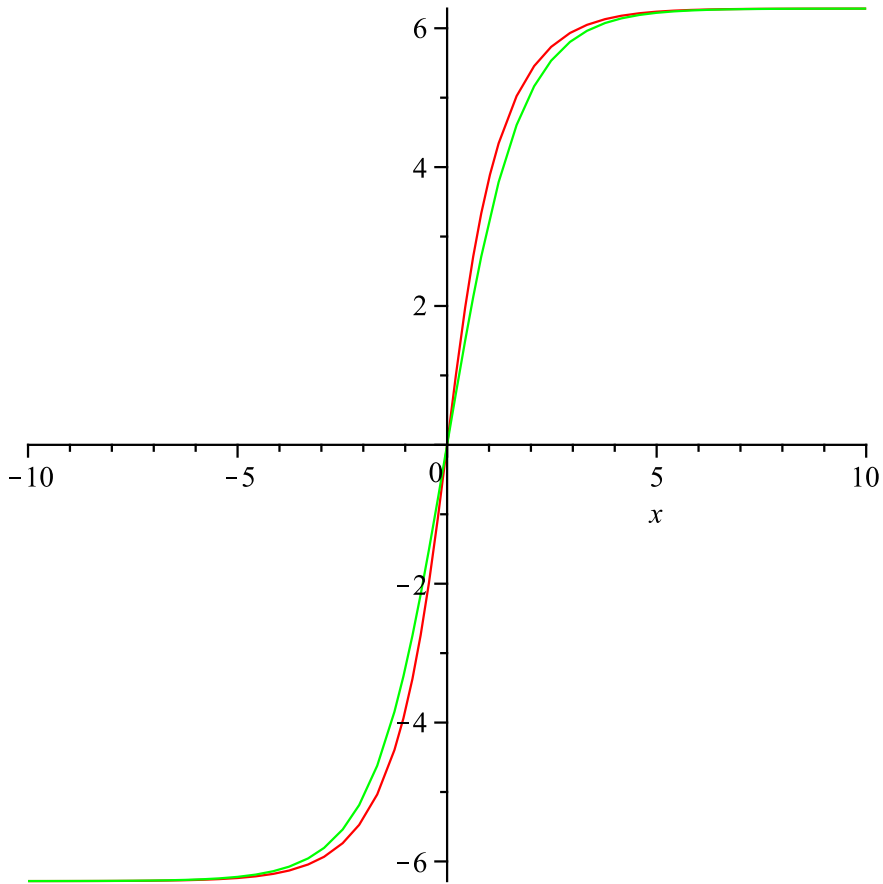}\end{includegraphics}
\caption{Kink fields given by equation (\ref{sut1}) (red) and by equation (\ref{sut2}) (green) for $a=0.5$.}
\label{sut}
\end{center}
\end{figure}

In order to derive the approximate dynamics, the task is therefore to compute the numerical integrals in (\ref{num in}).  In figure \ref{realdyn} we plot such approximate dynamics for $a_{0}=5$ and $u=0.3$.  We also include the dynamics according to a standard Runga-Kutta code.  We see that the approximate dynamics are qualitatively correct and are accurate before the ``apparent'' point of least approach. The motion according to (\ref{kinkansatz}) is seen to take place in a stronger potential than the actual potential.  This can be compared to Sutcliffe's analysis \cite{Sutcliffe:1993wc} for the sine-Gordon model where rather than using a patching of two kinks like we have done in this chapter, that is

\begin{equation}
\label{sut1}
\psi_{1}=4\arctan{e^{x-a}}+4\arctan{e^{x+a}}-2\pi,
\end{equation}
the following patching is used

\begin{equation}
\label{sut2}
\tan\left(\frac{\psi_{2}}{4}\right)=e^{x-a}-e^{-(x+a)}.
\end{equation}
In figure \ref{sut}, we plot $\psi_{1}$ and $\psi_{2}$ with the kinks close together ($a=0.5$).  We see that the kinks, according to (\ref{sut2}), start to deform slightly compared to (\ref{sut1}).  This has the effect of ``softening'' the ``apparent potential'' and is what gives the results presented in \cite{Sutcliffe:1993wc} where both dynamics remain more or less the same for all time.  The patching for $\phi^6$ kinks presented in this chapter does not, in the same way, take account of solitons deforming when they are close to each other.  However, the early time dynamics according to our approximation show that in this regime it works well numerically.  Such dynamics have also been calculated for a smaller initial speed ($u=0.1$) and results obtained are qualitatively the same.

%\begin{figure}[!htb]
%\begin{center}
%\begin{includegraphics}[width=12cm]{realdynamics.eps}\end{includegraphics}
%\caption{The ``real dynamics'' for $a_{0}=5$ and ${\dot a}(t=0)=-0.3$.}
%\label{realdyn}
%\end{center}
%\end{figure}

\section{Conclusion}
In this paper, we have investigated the collective coordinate approximation in the context of $\phi^6$ kinks with the aim of showing how to test the validity of the method, which is much used in the study of solitons.  We have found that the Bogomolny equations can be solved exactly for this model leading to 1-soliton solutions that can then be ``patched'' together.  This field led to a dynamical system for which the approximate motion was found via a series of numerical integrals.

The ``real dynamics'' were calculated through a numerical code and presented.  As expected, two kinks moving towards each other at non-relativistic speeds, slow down, because of their mutual repulsion, and eventually momentarily stop, then move apart.

With our initial conditions, the approximation we have made leads to dynamics which, for time before $t\approx 10,$ are of a good level of accuracy.  Due to the way we patch the kinks together, the solitons observe a potential that is stronger than the actual potential.  We conclude that the collective coordinate approximation presented in this paper leads to dynamics that are qualitatively the same as the actual dynamics and works well numerically in the early time regime, and that the approximation could be improved if the deformation of solitons as they begin to coalesce could be taken into account.

\section*{Acknowledgements}
The author would like to thank Steffen Krusch for discussions.  SWG also gratefully acknowledges the EPSRC and the SMSAS of the University of Kent for funding.

\begin{small}

\end{small}

\begin{thebibliography}{99}

\bibitem{Manton:2004tk}
  N.~S.~Manton and P.~Sutcliffe, {\it Topological solitons},
%\href{http://www.slac.stanford.edu/spires/find/hep/www?irn=6000355}{SPIRES
%  entry}
(Cambridge University Press, 2004).

\bibitem{Bogomolny}
  E.~B.~Bogomolny,
  ``The stability of classical solutions,''
  Sov.\ J.\ Nucl.\ Phys {\bf 24} (1976) 449.

\bibitem{Dorey:2011yw}
  P.~Dorey, K.~Mersh, T.~Romanczukiewicz and Y.~Shnir,
  ``Kink-antikink collisions in the $\phi^6$ model,''
  Phys.\ Rev.\ Lett.\  {\bf 107} (2011) 091602
  [arXiv:1101.5951 [hep-th]].

\bibitem{Manton:1981mp}
  N.~S.~Manton,
  ``A remark on the scattering of BPS monopoles,''
  Phys.\ Lett.\  B {\bf 110} (1982) 54.

\bibitem{Atiyah}
  M.~F.~Atiyah and N.~J.~Hitchin,
  {\it The Geometry and Dynamics of Magnetic Monopoles},
  (Princton Univ.~Press, Princeton, NJ, 1988).

\bibitem{Gibbons:1986df}
  G.~W.~Gibbons and N.~S.~Manton,
  ``Classical and quantum dynamics of BPS monopoles,''
  Nucl.\ Phys.\  B {\bf 274} (1986) 183.

\bibitem{Ruback:1988ba}
  P.~J.~Ruback,
  ``Vortex string motion in the Abelian Higgs model,''
  Nucl.\ Phys.\  B {\bf 296} (1988) 669.

\bibitem{Samols:1991ne}
  T.~M.~Samols,
  ``Vortex scattering,''
  Commun.\ Math.\ Phys.\  {\bf 145} (1992) 149.

\bibitem{Ward:1985ij}
  R.~S.~Ward,
  ``Slowly moving lumps in the Cp**1 model in (2+1)-dimensions,''
  Phys.\ Lett.\  B {\bf 158} (1985) 424.

\bibitem{Sutcliffe:1993wc}
  P.~M.~Sutcliffe,
  ``Classical and quantum kink scattering,''
  Nucl.\ Phys.\  B {\bf 393} (1993) 211.

\end{thebibliography}
\end{document}